\documentclass[twocolumn,showpacs,preprintnumbers,amsmath,amssymb,APSl,prd,nofootinbib,superscriptaddress]{revtex4-2}

\usepackage{dcolumn}
\usepackage{bm}
\usepackage{ifpdf}
\usepackage{hyperref}
\usepackage{soul} 
\usepackage{bm}
\usepackage[dvipsnames]{xcolor}
\usepackage{color,graphicx,graphics}
\usepackage[spanish,english]{babel}
\usepackage[latin1]{inputenc}
\usepackage[OT1]{fontenc}
\usepackage[justification=centering]{caption}
\usepackage{makeidx}
\usepackage{epsfig}
\usepackage{natbib}
\usepackage{epstopdf}
\usepackage{mathrsfs}
\hypersetup{colorlinks=true, linkcolor=blue, citecolor=green}
\usepackage{enumerate}
\usepackage{multirow}
\usepackage{subcaption}
\usepackage{booktabs}
\usepackage{amsmath}
\usepackage{mathtools}
\usepackage{orcidlink}

\DeclareCaptionFormat{myformat}{\justifying\noindent#1#2#3}
\usepackage{ragged2e}  

\definecolor{red}{rgb}{1,0,0}

\def\+{^\dagger}

\def\<{\leftarrow}
\def\>{\rightarrow}

\def\({\left(}
\def\){\right)}

\newcommand{\bi}{\begin{itemize}} 				\newcommand{\ei}{\end{itemize}}
\newcommand{\benu}{\begin{enumerate}} 		\newcommand{\enu}{\end{enumerate}}
\newcommand{\bd}{\begin{dinglist}{0}}     \newcommand{\ed}{\end{dinglist}}
\newcommand{\bfig}{\begin{figure}[htbp]}  \newcommand{\efig}{\end{figure}}
        			
\newcommand{\bc}{\begin{center}} 				  \newcommand{\ec}{\end{center}}
\newcommand{\be}{\begin{equation}} 				\newcommand{\ee}{\end{equation}}
\newcommand{\bsub}{\begin{subequations}}  \newcommand{\esub}{\end{subequations}}
\newcommand{\ba}[1]{\begin{array}{#1}} 		\newcommand{\ea}{\end{array}}
\newcommand{\bea}{\begin{eqnarray}}
\newcommand{\eea}{\end{eqnarray}}

\newcommand{\der}{\mathrm{d}}

\begin{document}
\title{Observational appearance and photon rings of non-singular black holes from anisotropic fluids}

\author{David D\'iaz-Guerra \orcidlink{0000-0003-0859-917X}} 
\email{ddiazgue@ucm.es}
\affiliation{Departamento de F\'isica Te\'orica, Universidad Complutense de Madrid, E-28040 Madrid, Spain}

\author{\'Angel Rinc\'on \orcidlink{0000-0001-8069-9162}}
\email{angel.rincon@physics.slu.cz}
\affiliation{
Research Centre for Theoretical Physics and Astrophysics, Institute of Physics, Silesian University in Opava,
Bezru\v{c}ovo n\'am\v{e}st\'i
13, CZ-74601 Opava, Czech Republic.
} 
\affiliation{
Instituto Universitario de Matem\'atica Pura y Aplicada, Universitat Polit\`ecnica de Val\`encia, Valencia 46022, Spain.
}

\author{Diego Rubiera-Garcia \orcidlink{0000-0003-3984-9864}} \email{drubiera@ucm.es}
\affiliation{Departamento de F\'isica Te\'orica and IPARCOS,
	Universidad Complutense de Madrid, E-28040 Madrid, Spain}

\date{\today}
\begin{abstract}

We consider the optical appearance of a non-singular, spherically symmetric black hole from 
Eddington-inspired Born-Infeld gravity coupled to anisotropic fluids. Such a black hole has a single (external) horizon located very near the Schwarzschild radius, $r_h=2M$, while its surface of unstable bound geodesics (photon sphere) is located at a moderately shortened radius than its Schwarzschild counterpart. Relying on a geometrically and optically thin accretion disk with a monochromatic emission described by suitable adaptations of Standard Unbound profiles previously employed in the literature, we generate images of this solution, which displays relevant modifications to the typical photon ring and central brightness depression features found in black hole images. In this sense, we fit the width of the two first photon rings in order to reconstruct the Lyapunov exponent of nearly-bound geodesics characterizing the theoretical ratio of successive rings. Such an exponent is tightly attached to observational features of photon rings such as their relative intensities in time-averaged images and the time-scale of hot-spots. Our results point out that non-singular black holes of this type are hard to distinguish from their Schwarzschild counterparts using this method alone, since the theoretical, numerical, disk-modeling, and observational uncertainties are too entangled with one another to allowing a neat distinction of such an exponent. It also points out to the need of incorporating dynamical settings such as hot-spots or quasi-normal modes from gravitational wave ringdowns as a way to circumvent such difficulties.

\end{abstract}

\maketitle

\section{Introduction}

Nowadays, the mounting observational evidence supporting the existence of black holes \cite{Bambi:2017iyh,Genzel:2024bh} has not only made them a prominent topic in contemporary astrophysics but also underpins their mathematical elegance, reinforcing their unique status as one of the simplest yet profound predictions of Einstein's General Relativity (GR). Indeed, the combination of the uniqueness theorems and the no-hair conjecture tells us that black holes are described by solely three parameters: mass, angular momentum, and electric charge \cite{Heusler:1996}. The astrophysical signatures of black holes are tightly linked to two surfaces harbored by them. On the one hand, trapped surfaces, i.e., two-dimensional space-like surfaces whose orthogonal  inward and outward future-directed null congruences are converging. On the other hand, the photon shell (the photon sphere in the spherically symmetric case), namely, the surface of unstable bound geodesics.  

Astrophysical observations have classified black holes into three main categories based on their mass: (i) stellar-mass $(4 - 100) M_{\odot}$, (ii) supermassive $(10^5 - 10^{10}) M_{\odot}$, and (iii) intermediate $(10^2 - 10^5) M_{\odot}$. Stellar-mass black holes, formed from the gravitational collapse of massive stars after a supernova phase, are detected through X-ray spectroscopy \cite{Bambi:2015kza} and gravitational wave signals in binary systems \cite{LIGOScientific:2016aoc}. Supermassive black holes, with still unclear origins, are investigated via the motion of S-stars in the Milky Way \cite{DellaMonica:2021fdr} and imaging of their accretion disks \cite{EventHorizonTelescope:2019dse,EventHorizonTelescope:2022wkp}. Finally, intermediate black holes have less conclusive evidence, relying on X-ray and gravitational wave data \cite{GreStrHo}, making their existence more debated. A fourth category is that of primordial black holes \cite{Garcia-Bellido:2017mdw,Escriva:2022duf}, formed at the onset of inflation and bounded by a mass $\gtrsim 10^{-18}M_{\odot}$, which is still only within the theoretical realm despite intensive observational searches \cite{Sasaki:2016jop,Garcia-Bellido:2017aan}.

Black holes, given the extreme environments they provide us with, also play a pivotal role in the limits of GR itself. In fact, while the near-horizon physics can be consistently associated to the astrophysical observations above, inside their innards black holes hoard several theoretical difficulties, namely, space-time singularities \cite{Senovilla:2014gza}, closed time-like curves \cite{Godel:1949ga}, mass inflation \cite{Poisson:1989zz}, and the information loss paradox \cite{Hawking:2005kf,Chen:2014jwq}. Space-time singularities are the unavoidable development of incomplete geodesic curves inside the event horizon. Closed time-like curves refers to particle's paths that return to the same location in time. Mass inflation is the unbounded growth of curvature, for local observers, as the inner horizon is approached. And the information paradox corresponds to the contradiction between the thermal nature of Hawking's emission and the unitary evolution demanded by the tenets of quantum mechanics.

The tension between the above theoretical difficulties displayed by canonical black holes and their reliability to successfully describe astrophysical phenomenology has compelled the community to search for an upgraded description of black holes which are devoid of some (or all) of the problems above. Focusing on space-time singularities, the so-called regular black holes are intended to provide a completion of canonical black holes via an everywhere geodesically complete description of every null and time-like trajectory. Furthermore, typically boundness of every curvature scalar made up of contractions of geometrical objects is also required. For a review of this topic see e.g. \cite{Bambi:2023try}. 

Given the strong connection provided by the singularity theorems between geodesic incompleteness and energy conditions, the natural place for regular black holes to exist in within modified theories of gravity. These theories are intended to supersede GR on its strong-field regime, while leaving current observational confirmations mostly untouched. While the new theory is widely believed to necessarily account for interacting quantum matter and gravity (i.e. a quantum theory of gravity), more modest approaches keep the idea of gravity as a manifestation of a space-time imbued with geometrical properties while dropping/amending some of its mathematical/physical principles, see e.g. \cite{DeFelice:2010aj,Clifton:2011jh,Nojiri:2017ncd}.

In a geometrical formulation of gravity, the metric defines local distances, while the connection governs parallel transport. These distinct concepts give rise to metric-affine theories of gravity based on curvature, torsion, and non-metricity \cite{Sotiriou:2006qn,Olmo:2011uz}. GR belongs to this family, via the lowest-order curvature scalar. Its torsion-based (teleparallel gravity) and non-metricity-based (symmetric teleparallel gravity) versions yield equivalent field equations, but differ in boundary terms \cite{BeltranJimenez:2019esp}. In curvature-based GR, the connection is given by the Levi-Civita one, defined in terms of the metric. In a general metric-affine formulation of gravity, however, the metric and connection are independent of each other, allowing for new actions with higher-order curvature scalars and distinct field equations.

The research in the literature has identified a sub-family of such theories  satisfying the following properties: (i) second-order field equations, (ii) absence of ghost-like instabilities; (iii) absence of additional propagating degrees of freedom; (iv) a well-defined Hamiltonian and boundary problem; (v) recovery of GR dynamics in vacuum; (vi) tractable-enough field equations to apply analytical/numerical methods. Such a family is dubbed as {\it Ricci-based gravity} \cite{Afonso:2018bpv}, and includes popular theories such as $f(R)$, quadratic gravity, or Eddington-inspired Born-Infeld (EiBI) gravity \cite{Banados:2010ix}. Furthermore, several methods have been developed in order to solve its corresponding field equations \cite{Afonso:2018mxn}.  Using these methods, solutions sourced by scalar, electromagnetic, and fluid fields have been found \cite{Olmo:2022rhf}, revealing a large variety of configurations in terms of horizons and regularity features, while having a rich phenomenology in astrophysical \cite{Shaikh:2019jfr,Nascimento:2018sir,Shao:2020weq,Bora:2022qwe,Wojnar:2022ttc,Soares:2023err,Zeng:2025xoe,Magalhaes:2023har}, cosmological \cite{Bauer:2008zj,Gialamas:2019nly,Enckell:2018hmo,Rubio:2019ypq,Gialamas:2021enw,Gomes:2023xzk} and particle physics \cite{Barker:2024ydb}.

The main goal of this work is to find the optical appearance of a non-singular black hole solution within the framework of a Ricci-based gravity of the EiBI type (for a review of this theory see \cite{BeltranJimenez:2017doy}) supported by an anisotropic fluid. This field of study is of current interest under the light of the observations made by the Event Horizon Telescope (EHT) Collaboration regarding the imaging of the supermassive objects at the heart of the M87 
\cite{EventHorizonTelescope:2019dse} and  Milky Way \cite{EventHorizonTelescope:2022wkp} galaxies. This new opportunity has motivated the community to search for images of both alternative black holes \cite{Hou:2021okc,Afrin:2021wlj,Kuang:2022ojj,Daas:2022iid,Meng:2023htc,Xu:2025iwg,Glampedakis:2021oie} and horizonless ultra-compact objects \cite{Olivares:2018abq,Joshi:2020tlq,deSa:2024dhj,Fauzi:2025rcc,Zhao:2025yhy,deSa:2025nsx}. 

The EiBI theory smoothly recovers, on its lowest-orders, the Einstein-Hilbert action of GR (with a cosmological constant term) plus quadratic curvature corrections. On the other hand, the anisotropic fluid assumption, whose formulation in our framework naturally verifies the energy conditions, captures many cases of matter sources of physical interest \cite{Daouda:2012nj,Maurya:2017non,Kumar:2017tdw}, including non-linear electrodynamics. Such a combination of gravity and matter sources gives rise to several solutions that restore the geodesically complete character of the background geometry, hence removing the issue with space-time singularities which plague GR black holes. By picking one of them, we generate its optical images when illuminated by a thin accretion disk, which display the typical structure of a bright annular region surrounding a central dark region. Furthermore, the bright region is decomposed into a sequence of infinite photon rings, whose features we analyze in detail using the full width at half maximum method. The information found this way can then be compared with the GR-based canonical black holes and  the current capabilities by the EHT images as well as with those of the planned very-long baseline interferometric devices such as the ngEHT \cite{Tiede:2022grp} and the Black Hole Explorer \cite{Johnson:2024ttr} in searching for traces of new gravitational physics.

This work is organized as follows. In Sec. \ref{S:II} we establish our line element and discuss its most salient physical features from a theoretical perspective. In Sec. \ref{S:III} we develop the formalism for generation of optical images, based on the null geodesic equation, and set our choice of parameters for the gravity plus matter combination. In Sec. \ref{S:IV} we generate full images of our geometry for a geometrically and optically thin disk emitting according to three suitable adaptations of the Standard Unbound distribution previously employed in the literature to match specific scenarios of GRMHD simulations. An analysis of the structure of the photon rings of such images according to their width ratio is carried out in Sec. \ref{S:V} due to their potential observability, while in Sec. \ref{S:VI} we gather the conclusions of our paper and some discussion. We work in units $\kappa^2=\frac{8\pi G}{c^4}=1$.

\section{Non-singular black holes in EiBI gravity} \label{S:II}

\subsection{Presentation of the geometry}

The solutions we are considering in this work arise from a combination of EiBI gravity, characterized by a certain length scale $l_{\epsilon}^2=\vert \epsilon \vert$ (with $\epsilon$ the EiBI parameter appearing the action) plus a matter sector composed of an anisotropic fluid characterized by a matter scale $l_{\beta}^2=\beta^4/2$ (with $\beta$ the parameter appearing in the corresponding energy-momentum tensor), which were first found in \cite{Menchon:2017qed}. A recap of the derivation of our setting extracted from such a reference can be found in Appendix \ref{App1}. In Schwarzschild-like coordinates $(t,x,\theta,\varphi)$ the line element is best presented using dimensionless variables $x \to x/r_c$ and a radial function $z(x)=r/r_c$ (encoding the areal radius of the two-spheres as $S=4\pi r_c^2 z^2(x)$) with $r_c \equiv \vert \beta \vert$ and reads as follows
\begin{equation} \label{eq:lineel}
ds^2=-A(x) dt^2+\frac{dx^2}{\Omega_1^2(x) A(x)}+z^2(x) d\Omega^2,
\end{equation}
with the following definitions and conventions: 
\begin{eqnarray}
A(x)&=& \frac{1}{\Omega_1} \left(1-\frac{r_S(1+\delta_1 G(z))}{r_c z \Omega_2^{1/2} } \right),   \\
\frac{dG}{dz}&=& \frac{z^2  \Omega_1}{(z^4-1)\Omega_2^{1/2}}, \\
\Omega_1&=&1+\frac{\lambda^2 (z^4+1)}{(z^4-1)^2}, \\
\Omega_2 &=&1 - \frac{\lambda^2 }{z^4-1}.
\end{eqnarray}
Here $r_S=2M$ is the usual Schwarzschild's radius with $M$ the total space-time, asymptotic mass (arising as usual as an integration constant out of the field equations), $\lambda^2 \equiv l_{\epsilon}^2 / l_{\beta}^2$ is the ratio between the scales defining the gravity and matter sectors,  and the constant
\begin{equation}
    \delta_1=\frac{2r_c^3}{r_S \beta^4} = \frac{2}{r_S \vert \beta \vert}
\end{equation}
is a composition of all these constants, playing a key role in the global structure of the solutions. The metric function $G(z)$ can be integrated analytically, and after some suitable manipulations the result can be re-arranged as 
\begin{align}
    G(z) &= - \frac{2}{15 \lambda^2 z} \Big(-5(1+2\lambda^2) {F}_1 \left(\frac{1}{4},\frac{1}{2},\frac{1}{2},\frac{5}{4};\frac{1}{z^4},\frac{1+\lambda^2}{z^4} \right) \nonumber \\
    &+ \frac{1+\lambda^4}{z^4} F_1 \left( \frac{5}{2},\frac{1}{2},\frac{1}{2},\frac{9}{4};\frac{1}{z^4},\frac{1+\lambda^2}{z^4} \right) \Big) \nonumber 
    \\
    &+ \frac{5z^3(-2-\lambda^2 + 2z^{-4})(1+\lambda^2-z^4)^{1/2}}{15 \lambda^2 (1-z^4)^{3/2}}, \label{eq:GZ}
\end{align}
where $ {F}_1$ is an hypergeometric function.

The final ingredient in our line element is the radial function $z(x)$, in terms of which the areal radius of the two-spheres is expressed as $S=4\pi r_c^2 z^2(x)$. This function is implicitly defined via the relation\footnote{It should be stressed that in metric-affine theories of gravity of the kind considered here the field equations can be typically cast in standard-GR like form in terms of an auxiliary metric whose radial coordinate $x$ is related to the radial function of the space-time metric $z^2(x)$ as $x^2=z^2(x)\Omega_{2}$ (see e.g. \cite{Olmo:2012nx}). Taking a derivative upon this relation and after some manipulations, one arrives to the expression (\ref{eq:dzdx}).}
%
%
\begin{equation} \label{eq:dzdx}
    \frac{dx}{dz}=\frac{\Omega_1}{\Omega_2^{1/2}},
\end{equation}
and while it can be analytically integrated, the full expression is very cumbersome and writing it explicitly adds little to our purposes. Nonetheless, its relevance lies on the fact that at the location $x=0$, corresponding to $z=z_c =(1+\lambda^2)^{1/4}$,  the radial function attains a minimum.

\subsection{Physical interpretation}

For large distances, the line element captures a black hole geometry that behaves as 
\begin{equation}
g_{tt} \approx - g_{rr}^{-1} \approx - \left(1 - \frac{r_S}{r} + \frac{2}{r^2} -\frac{2l_{\epsilon}^2}{r^4} + \mathcal{O} \left(\frac{1}{r^5} \right) \right),
\end{equation}
and therefore it has, on its three first terms, the same functional dependence as a Reissner-Nordstr\"om solution of GR, with additional correction in the gravity theory's parameter. This resemblance is exact (in this limit) provided that we identify the charge with the parameters of our solution as $Q =2r_c^4/\beta^4=2$. 
Note that several theories beyond GR predict gravitational potential corrections of the form $\propto r^{-4}$, often inspired by gravitational frameworks incorporating quantum effects into space-time. An example can be found in \cite{Lewandowski:2022zce}, where the quantum Oppenheimer-Snyder model in loop quantum cosmology is employed to develop a novel quantum black hole model. The resulting metric tensor is a deformed Schwarzschild solution, reflecting quantum effects that modify classical geometry. Similarly, in \cite{Junior:2023ixh}, new regular black hole solutions were found in conformal Killing gravity coupled with nonlinear electrodynamics. These corrections signify a quantum imprint of new physics, highlighting the interplay between GR and quantum mechanics in black hole descriptions. Furthermore, the mass term in our case has the right sign and can be identified with the ADM mass (since EiBI gravity does not modify the gravitational dynamics outside the matter sources), this way bypassing the typical problem with higher-order curvature theories in which the presence of additional fields makes this identification impossible,  something avoided in these theories due to the fact that no additional fields are propagated in this gravitational theory.

Significantly, the central behavior is neatly different. This is mainly due to the development of a bounce in the radial coordinate, as given by the existence of a minimum $r=r_c$ at $x=0$, as follows from Eq.(\ref{eq:dzdx}). Such a bounce prevents the focusing of geodesics ascribed to the singularity theorems, allowing for the propagation of null and time-like geodesics to the region $x<0$ (as explicitly shown in \cite{Menchon:2017qed}) while having everywhere finite values of the areas of the two spheres $S \equiv 4\pi r^2(x) \geq 4\pi r_c^2$. As a side token, for the solution considered here the energy density of the fluid remains everywhere finite (as proven in  \cite{Menchon:2017qed}) and, while the curvature scalars generically diverge at the bounce location, they remain finite in the special case $\delta_1=\delta_c$, where $\delta_c$ is a constant arising out of the integration of the field equations (see \cite{Menchon:2017qed} for its explicit expression). Furthermore, divergences in such curvature scalars do not necessarily entail entirely destructive forces upon time-like observers, as analyzed in \cite{Olmo:2016fuc} for EiBI gravity with electromagnetic fields.

On the other hand, the changes to the inner region of the solutions have also an impact in their global structure in terms of horizons. In this sense, the configurations with $\delta_1 > \delta_c$ resemble the Reissner-Nordstr\"om solutions in the sense of having two, one (degenerate) or none horizons. On the other hand, those with $\delta_1 < \delta_c$ are Schwarzschild-like with a single non-degenerate horizon. And finally, those special cases with $\delta_1=\delta_c$ feature either one (non-degenerate) or none horizons, depending on the values of the remaining parameters. 

The bottom line of the above discussion is that these configurations can be consistently interpreted as {\it wormholes}, namely, structures connecting distant regions of the space-time. Furthermore, they represent either non-traversable wormholes, akin to black holes with at least an external horizon but a regular interior; or traversable wormholes, namely, a kind of horizonless object allowing the transit of matter and radiation from both sides of them. Both of them are non-singular, geodesically-complete space-times, filled everywhere with an anisotropic fluid, though far from the wormhole the decaying behavior of the fluid make the solution to behave as a Reissner-Nordstr\"om geometry.

\section{Quantities for optical appearances} \label{S:III}

\subsection{Equations of null geodesics}

The optical appearance of a black hole is dominated by two main features: a bright annular region (the photon ring) and a dark central region (the shadow). The ring appears in the EHT observations \cite{EventHorizonTelescope:2019dse,EventHorizonTelescope:2022wkp} as unresolved and asymmetric in brightness (due to Doppler and relativistic beaming effects), while the shadow is inferred below a certain given brightness threshold (corresponding to a $\sim 10\%$ of the top intensity). Theoretically, the origin of both these two features can be tracked to the existence of unstable null bound surfaces (the photon sphere for the present spherically symmetric case) rather than to event horizons. Such surfaces correspond to photon geodesic trajectories that, when traced backwards from the observer's screen (using the time-reversal symmetry of the geodesic equations), asymptote to the maxima of the effective potential for null geodesics. We shall assume, without any loss of generality, that the photons' equation of motion are solved at the equatorial plane $\theta = \pi/2$, and then, thanks to the spherical symmetry, rotate the trajectory to be on any $\theta$-plane. Note that this is a convenient gauge/coordinate choice rather than a restriction on the regions occupied by photon's orbits. Then, the effective potential is given by 
\begin{equation}
    V(r)=\frac{A(r)}{r^2},
\end{equation}
corresponding to the equation of null geodesics motion which in the present case is written as (see \cite{Olmo:2023lil} for details of the corresponding derivation)
\begin{equation} \label{eq:geoeq}
   \frac{1}{\Omega_1^2} \dot{x}^2=\frac{1}{b^2}-V(r),
\end{equation}
where an overdot denotes a derivative with respect to the affine parameter, while $b \equiv L/E$ is the impact factor, defined as the ratio between the photon's angular momentum, $L=r^2 \dot{\phi}$, and the energy
$E=-A \dot{t}$, both conserved quantities in the asymptotically-flat, spherically symmetric line element (\ref{eq:lineel}).

The requirement of the critical points of the effective potential provides the relations
\begin{equation}
r_{ps}A'_{ps} - 2 A_{ps} =0 \quad ;
\quad 
b_c=\frac{r_{ps}}{\sqrt{A_{ps}}},
\end{equation}
where $r_{ps}$ stands for the photon sphere radius. We denote $A_{ps} \equiv A({r_{ps}})$, and $b_c$ is the critical impact parameter associated to light rays which asymptote to the photon sphere (dubbed as the {\it critical} impact parameter). Note that critical points may correspond to either maxima or minima of the potential; the latter are stable points and are dubbed as {\it anti-photon} spheres, and are of relevance for horizonless compact objects. It is important to stress that black holes must necessarily have photon spheres \cite{Carballo-Rubio:2024uas}, while for horizonless compact objects they always come (if present) in pairs: (at least) a photon sphere and anti-photon sphere together \cite{Cunha:2017qtt}.

At this point, and for the sake of the ray-tracing procedure, in which we need to find the deflection angle as a function of the radius, we can resort to an integration of the geodesic equation in terms either of the radial coordinate $x$ or of the radial function $r$, related via Eq.(\ref{eq:dzdx}). In terms of the former and using the conservation of the angular momentum, the geodesic equation reads as 
\begin{equation}
    \frac{d\phi}{dx}= \mp \frac{b}{\Omega_{1} r^2(x)} \frac{1}{\sqrt{1-\frac{Ab^2}{r^2(x)}}},
 \end{equation}
where $\mp$ for ingoing/outgoing geodesics. In this system the coordinate $x \in (-\infty,+\infty)$, with $x=0$ representing trajectories that would cross the wormhole throat. In terms of the latter system of coordinates, one can implement the change (\ref{eq:dzdx}), so in these new coordinates the line element restores its usual Schwarzschild-like shape (though in such a case one always needs to add the condition $r \geq r_c$ to account for the bounce). This way, the geodesic equation becomes, in these coordinates, as
\begin{equation}
    \frac{d\phi}{dr}= \mp \frac{b}{\Omega_{2}^{1/2} r^2} \frac{1}{\sqrt{1-\frac{Ab^2}{r^2}}}. \label{eq:geodesic}
\end{equation}
Note that, upon integration, each of the corresponding integrals for $\phi$ diverge at $r=r_{ps}$ for light rays with $b=b_c$. This condition corresponds to the infinite deflection angle expected for those photons with critical impact parameter.

In the sequel, we shall use this last form of the geodesic equation to carry out our integrations, due to its greater efficiency when running numerical simulations. To this end we shall un-do, in the expression of the line element and the metric components, the dimensionless variables $z(x)$, which are convenient to discuss the non-singular character of the space-time, and switch to dimensionfull ones, $r(x)$, which are more suitable to compare the imaging of these configurations with the one of the Schwarzschild black hole. Before going that way, however, we need to further characterize nearly-bound trajectories given their impact in observable features of black hole images. 

\subsection{(In)stability of null geodesics: the Lyapunov exponent} \label{sec:lya}

The trajectories of light rays around the non-singular black hole are governed by the geodesic equation \eqref{eq:geodesic}. The solutions to this equation (given an impact parameter $b$) relate the radial distance as a function of the deflection angle, i.e., $r(\phi;b)$. Using the time-reversal symmetry of the geodesic equations, one can perform a backwards ray-tracing procedure of those light rays arriving to the observer's plane image towards its source so that a given geodesic turn some deflection angle  either finding on their trajectory a turning point at some radial distance $r_{tp}$ (for $b>b_c$) or continuing its path until crossing the black hole event horizon $r_h$ (for $b<b_c$). Of particular interest are those trajectories that pass nearby the photon sphere, $r_\text{ps}$ (i.e., those with $b \gtrsim b_c$), and which are characterized by their inherently unstable character. To analyze such an instability, we consider a light ray starting its trip at a radial location which deviates by a small amount $\delta r$ from the photon sphere, that is, $r_o = r_\text{ps} + \delta r$, with $\delta r \ll r_{ps}$. Upon perturbation of the geodesic equation (\ref{eq:geoeq}), the first order vanishes at the photon sphere (due to the critical character of the effective potential there), while at second order we get the equation
\begin{equation} \label{eq:lyaeq}
    \pi \frac{\der \delta r}{\der \phi} = \gamma_{ps} \, \delta r,
\end{equation}
where the factor $\pi$ is added to normalize the turns after integrating one complete half-orbit around the black hole, and $\gamma_{ps}$ is the Lyapunov lensing exponent. For a general spherically symmetric metric of the form
\begin{equation}
    ds^2 = - A(r) dt^2 + B(r) dr^2 + C(r) d\Omega^2,
\end{equation}
the Lyapunov exponent can be shown to be given by (see \cite{Kocherlakota:2023qgo,Kocherlakota:2024hyq} for a general discussion on critical exponents of nearly-bound geodesics)
\begin{equation}
    \gamma_{ps} = \pi \sqrt{ \frac{1}{2} \bigg \vert \frac{(A C'' - C A'')}{AB} \bigg \vert}_{r_{ps}} .
\end{equation}
For the geometry considered in the present work, as given by the line element (\ref{eq:lineel}), this exponent reads explicitly (in terms of the radial coordinate $r$) as
\begin{equation}
    \gamma_{ps}= \pi \vert \Omega_1 \vert \sqrt{ \bigg \vert \left( A -  \frac{r^2 A''}{2} \right) \bigg \vert}_{r_{ps}} . \label{eq:lyapunov}
\end{equation}
From Eq.(\ref{eq:lyaeq}) one finds that the photon's radial location, after a deflection angle $\phi$ has elapsed, is given by
\begin{equation*}
    r=r_0  e^{\frac{\gamma_{ps} \phi}{\pi}}.
\end{equation*}
According to this expression, the Lyapunov lensing exponent can be understood a measure of the instability scale  associated to nearly-bound geodesics, in this case expressed as the drift of the radial function after a deflection angle $\phi$. For observational purposes, however, it is better to cast the equation above in terms of the number of half-turns performed by the photon around the black hole, defined as $n \equiv \tfrac{\phi}{\pi}$. In fact, photons are said to belong to the $n^{th}$ lensing band (with $n \geq 1$) if and only if they intersect the equatorial plane exactly $(n+1)$ times on its path from the source to the observer \cite{Cardenas-Avendano:2023obg}. This way, photons in the $n^{th}$ lensing band (i.e. which have executed on their winding a  number $n$ of half-turns around the black hole) will be identified with the creation (in the optically-thin limit) of a set of images on the observer's screen, which are dubbed as the {\it photon rings}. Such a sequence of photon rings can be understood as strongly-lensed images of the disk, dominated by emission for small radii and subject to large gravitational redshift, and governed on its behavior by the Lyapunov exponent above. In the limit $n \to \infty$, this sequence of rings approaches a {\it critical curve} in the observer's image plane, and which corresponds to the projection of the photon sphere there. Furthermore, this curve defines the outer edge of the black hole {\it shadow} since trajectories inner to it would have hit, on the ray-tracing procedure, the event horizon. However, in the physical scenario in which the main source of illumination is provided by an accretion disk, the actual central brightness deficit can be significantly reduced. For such realistic models, the inner emission (beyond the ISCO) that comes from plunging material following non-circular falling trajectories, contributes to the inner part of the image.

The Lyapunov lensing exponent, $\gamma_{ps}$, can be related to the Lyapunov time, $t_{ps}$, using the conserved equations of null geodesics as \cite{Kocherlakota:2024hyq}
\begin{equation}
    t_{ps} = \pi \frac{b_{ps}}{\gamma_{ps}} .\label{eq:lyapunov_time}
\end{equation}
The Lyapunov time characterizes the  instability time-scale of nearly-bound geodesics, and it is relevant for dynamical observations, such as hot-spots. These are local regions of the disk that are significantly brighter and hotter than their surroundings, supposedly induced by magnetic reconnection events \cite{Ripperda:2020bpz}. Furthermore, the Lyapunov time is also related to the decay time of test-field perturbations in the eikonal limit, and therefore connected to the imaginary frequency of the quasinormal modes of oscillations in such limit as \cite{Cardoso:2008bp,Pedrotti:2025idg}
\begin{equation} \label{eq:corre}
    \omega_I = \left(n + \frac{1}{2}\right) t_{ps}^{-1}.
\end{equation}
Due to its relation to both time-averaged and transient processes, the characterization of the Lyapunov exponents of nearly-bound geodesics will become a main goal of our subsequent analysis. 

\subsection{Choice of parameters} \label{sec:cop}

As stated in the previous section, the non-singular black holes considered in this work support a wide array of configurations. In order to pick one as representative as possible, we select the parameters $\lambda=2$ and $\beta = 1.33681$. These parameters are chosen in order for the non-singular black hole to have the same structure as the Schwarzschild one: a single non-degenerate horizon enclosing a internal space-like region. In this way, we avoid any issue with mass inflation phenomena that typically plague black holes with an interior horizon \cite{Carballo-Rubio:2022kad}. In addition, for this choice of parameters, the horizon is located very approximately at the Schwarzschild radius, $r_h \approx 2M$, this way allowing us to compare the imaging of both geometries in as a similar setting as possible.

\begin{figure}[t!]
\includegraphics[width=0.5\textwidth]{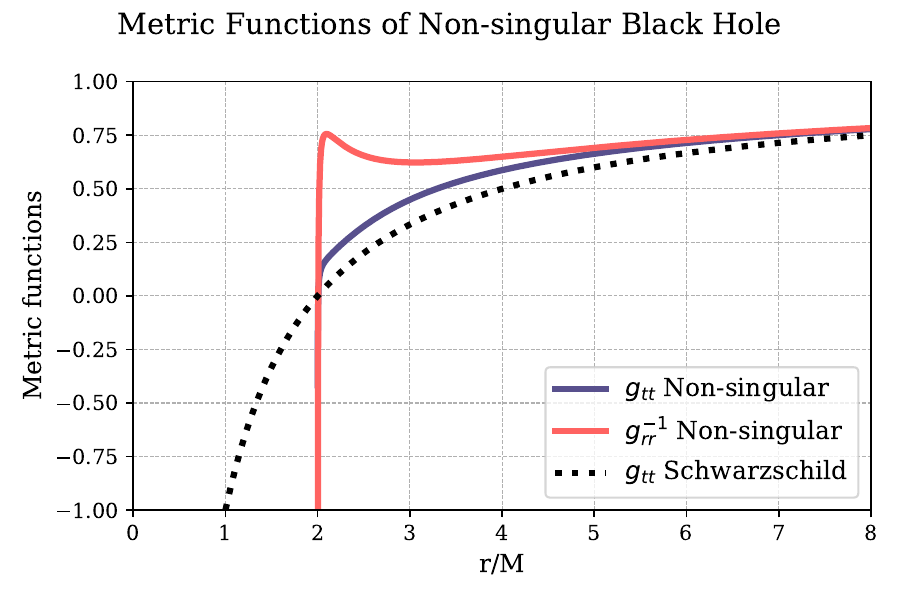}
\caption{Behavior of the metric components $g_{tt}$ (blue) and $g_{rr}^{-1}$ (red) as a function of $r/M$ for the EiBI non-singular black hole with the choice of parameters of Sec. \ref{sec:cop} as compared to the Schwarzschild one (dashed black).}
\label{fig:metricomp}
\end{figure}

In Fig. \ref{fig:metricomp}, we display the behavior of the metric components $g_{tt}(r)$ (blue) and $g_{rr}^{-1}(r)$ (red) for the non-singular black hole configurations, using the Schwarzschild-like coordinates ($t,r,\theta,\varphi)$. Additionally, we plot the same metric components of the Schwarzschild black hole, for which $g_{tt}=g_{rr}^{-1}$ (dashed black). As it can be neatly seen there, the horizon's location is pretty much the same in both cases, and the non-singular black hole asymptote to the Schwarzschild space-time at large distances, but the central behaviour is different. In the Schwarzschild case, the metric goes all the way down to $r=0$ until meeting there the incompleteness of all paths and the vanishing of the area of the two-spheres, $S=4\pi r^2 \to 0$. In the non-singular case, the metric components go also to $-\infty$ but now at a finite surface of radius $r=r_c(1+\lambda^2)^{1/4} \approx 1.999$ (in units of $M=1$), corresponding to the bounce location. Given the fact that the area of the two spheres $S=4\pi r^2(x)$ is finite, and furthermore that the metric components can be smoothly extended to the region $x<0$ (where $r >r_c(1+\lambda^2)^{1/4}$), the observer's path is well defined everywhere.

Moreover, the external region to the event horizon, which from the point of view of optical appearances is the only accessible one, is neatly different in the EiBI non-singular black holes and the Schwarzschild case. We therefore shall also expect non-negligible differences regarding their photon ring and shadow features. Indeed, the relevant parameters for the photon sphere are significantly different:
\begin{eqnarray}
    r_{ps} &\approx& 2.628M \quad (r_{ps}^S=3M) ,\\
    b_{ps} &\approx& 4.366M \quad (b_{ps}^S=3\sqrt{3M} \approx 5.196M) ,
\end{eqnarray}
where $S$ denotes the Schwarzschild values. It should be stressed that the value for the critical impact parameter for this choice of parameters is within the bounds established by the EHT Collaboration regarding the inferred, calibrated size of the shadow of the M87 and Sgr A$^*$ observations \cite{EventHorizonTelescope:2022xqj} (for a discussion of this result and its latent power to constraint black hole metrics see \cite{Vagnozzi:2022moj}). Therefore, this scenario provides significant but mildly enough modifications with respect to GR images so as to pass current tests while at the same time capable of providing observable differences in the structure of their photon rings, as we shall see in the sequel. 

The location of the photon sphere allows us to characterize the instability scale of circular null orbits. The lensing Lyapunov exponent ($\gamma_{ps}$), which measures the rate of exponential divergence of nearby geodesics in azimuthal angle (or, equivalently, in terms of the number of half-orbits $n$), and the corresponding Lyapunov time ($t_{ps}$), which sets the timescale for this instability, can be calculated by substituting the photon sphere's location into Eqs. \eqref{eq:lyapunov} and \eqref{eq:lyapunov_time}, which yields the results
\begin{eqnarray}
    \gamma_{ps}&\approx& 2.972 \quad (\gamma_{ps}^S=\pi \approx 3.141) \label{eq:lyapunov_th} , \\
    t_\text{ps} &\approx& 4.651M\quad (t_{ps}^S=3\sqrt{3}M \approx 5.196M ) \label{eq:lyapunov_thh} .
\end{eqnarray}
Both values are slightly smaller than their counterparts around the Schwarzschild spacetime. This shows that the instabilities decay faster in an EiBI non-singular black hole than in the standard Schwarzschild black hole.

\section{Optical appearances} \label{S:IV}

\subsection{Ray-tracing}

We perform backward ray-tracing by integrating the null geodesic equation in the space-time given by Eq.\eqref{eq:lineel}, which describes a static, spherically symmetric non-singular black hole within EiBI gravity using the Palatini approach, setting
the parameters described in Sec. \ref{sec:cop}.

The null geodesic equation \eqref{eq:geodesic} is numerically integrated (see Appendix \ref{App2} for an overview on how we handle the special functions appearing in the space-time metric functions, recall Eq.(\ref{eq:GZ})) from a large radial distance, $r_o \simeq 100 M$, where the space-time is approximately flat.  Each ray originates from a point on the plane specified by Bardeen's coordinates $(\alpha, \beta)$. These coordinates are interpreted as the polar coordinates in the observer's plane image.  We define a set of polar coordinates on the screen $(\rho, \vartheta)$, which due to the spherical symmetry we can associate directly with the impact parameter, with $\rho = b$.  By integrating from the observer to the black hole (using the sign $(-)$ in Eq.\eqref{eq:geodesic}), we determine whether a given light falls into the black hole or escapes back to infinity. The latter will happen whenever the right-hand side of the geodesic equation \eqref{eq:geodesic}  is zero, corresponding to a turning point, from where it is integrated again back to a large distance, using now the sign $(+)$. We use a fourth-order adaptive step-size Runge-Kutta method with fifth-order error checking to perform the integration. This ensures stability near the photon sphere, where the rays are susceptible to instabilities leading to diverging trajectories.

\begin{figure}[t!]\includegraphics[width=0.5\textwidth]{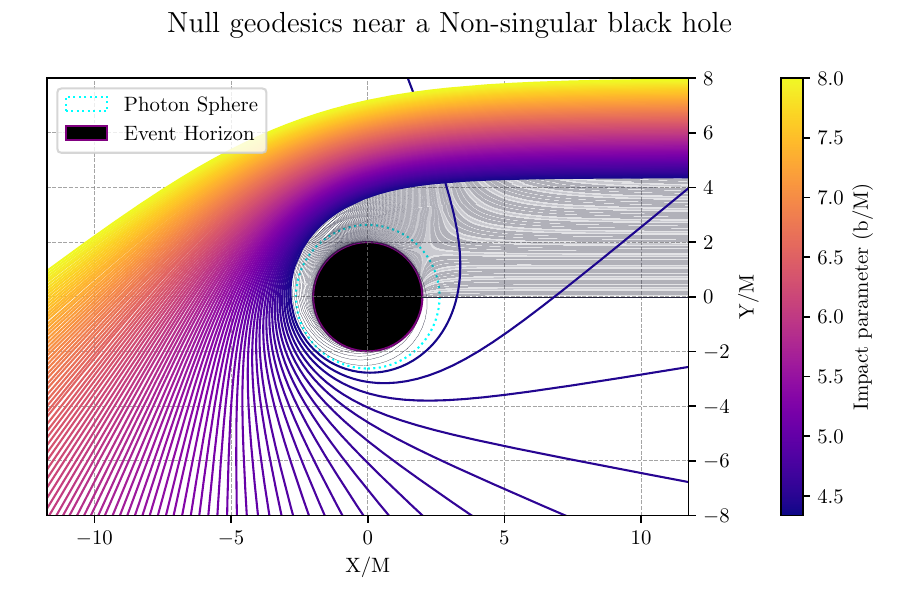}
    \caption{Ray-tracing of null geodesics from a screen at $r_o=100M$ to a non-singular EiBI black hole with the choice of parameters of Sec. \ref{sec:cop}. We draw the rays that correspond to the shadow, i.e. those corresponding to $b<b_{ps}\simeq 4.366$, in shade of gray. The rest of the rays find a turning point and go back to infinity, and are colored using the intrinsic impact parameter of the trajectory.}
    \label{fig:raytrace}
\end{figure}

In Fig. \ref{fig:raytrace}, we plot the rays emitted from the equator plane by an observer at infinity. We draw in a shade of gray the rays that fall into the black hole that correspond to the shadow, as defined by the region inner to the critical curve, that is, $b<b_{ps}$. The rest correspond to impact parameters $b>b_{ps}$ (depicted in different colors according to their closenss to the critical impact parameter), which return to infinity as they do not cross through the photon sphere at $r_{ph}$ but instead find a turning point around the black hole.

\subsection{Optically and geometrically thin accretion disk} 

We consider the optical appearance of an accretion disk orbiting a non-singular black hole described in the last section. The Event Horizon Telescope observations \cite{EventHorizonTelescope:2019pgp} suggest that realistic disks around supermassive black holes are optically thin but geometrically thick, extending a finite distance from the equator. 
To simplify the calculations, we employ the equatorial approximation \cite{Gralla:2020srx}, which assumes that the dominant emission originates from the equatorial plane, neglecting contributions from latitudes away from it. This assumption leads to a mild quantitative disagreement with several observables under the presence of magnetic fields \cite{Saurabh:2025kwb} and, in particular, with the values of the intensities of the photon rings, though not necessarily on the  relative intensity of the successive rings. Furthermore, the former deviation from the neglected effect of geometrical thickness can be corrected via the introduction of a ``fudge" factor  \cite{Cardenas-Avendano:2023dzo} to account for the otherwise ignored thickness of the disk, this way improving the agreement with time-averaged images of the photon rings \cite{Vincent:2022fwj}.

For the sake of this and to enhance our conceptual understanding of the underlying physics, we employ a simplified accretion disk description in place of more realistic and sophisticated models. Two concrete assumptions are adopted: (i) the disk comprises static emitters following circular geodesics outside the photon sphere, and (ii) inside this radius, the emitters follow plunging-like infall paths until crossing the event horizon. Obviously, this is a toy-like model whose goal is to highlight the contribution of photon propagating effects towards the generation of the image, neglecting other processes and effects that are required for a fully-relativistic emission models incorporating all physical ingredients known to play a role. 
While this is an usual approximation to the actual images, as has been regularly employed in the literature, we recognize that a full qualitative study requires a more refined treatment. Thus, while the inclusion of more realistic ingredients as radiative processes and proper radiative transfer, or the use of a mass accretion rate derived from observational fits, falls outside the goals of the current study, it is essential to acknowledge that accounting for these factors would likely lead to markedly improved model predictions.

\subsection{Transfer functions}

The ray-tracing procedure establishes a mapping between an observer at infinity and points on the accretion disk, such that for each ray that originates on the observer screen, we determine the intersection point with the disk, lying on the equator. The transfer functions, $r_n$, determine the location of the intersection of a given ray with impact parameter $b$, with the disk after $n$-half turns over the black hole (in this notation $n=0$ corresponds to the disk's direct emission, and $n=1,2,\ldots$ to the photon rings). 

\begin{figure}[t!]
    \includegraphics[width=0.46\textwidth]{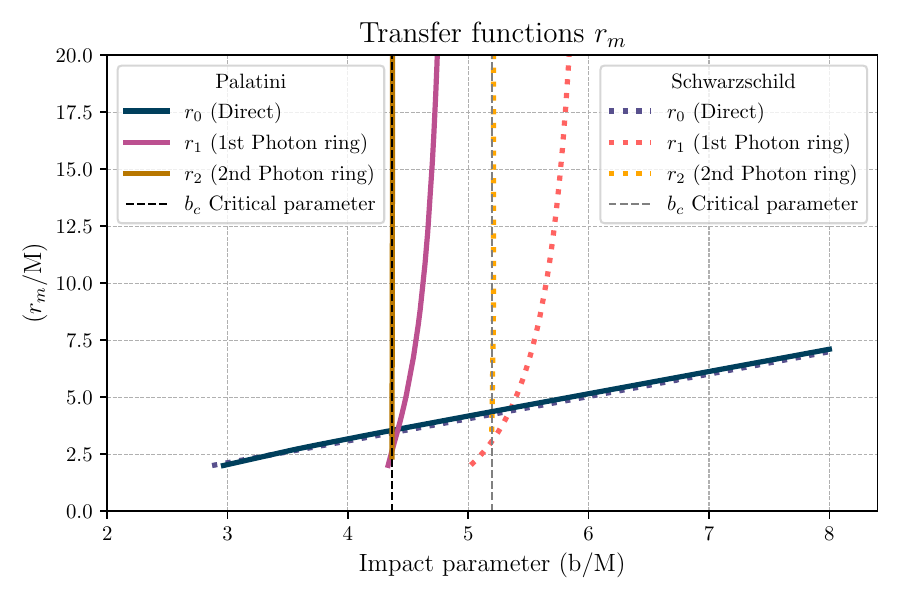}
        \caption{Transfer functions $r_n$ for $n=0,1,2$, which correspond to the disk's direct image and the first and second photon ring emissions, respectively, for the non-singular EiBI  black hole and the Schwarzschild solution. We plot as a vertical line the theoretical position of the shadow as determined by the respective critical impact parameters.}
    \label{fig:transfer}
\end{figure}

In Fig. \ref{fig:transfer}, we plot the transfer function for $n\leq 2$ for the non-singular EiBI black hole as well as the Schwarzschild black hole for comparison. The first transfer function $r_0$ corresponds to the direct emission of the disk, and we do not find any considerable difference with the Schwarzschild black hole. This is in agreement with the realization in the literature that the direct's emission is mainly determined by the properties of the accretion disk rather than the background geometry \cite{Lara:2021zth}. Meanwhile, the sequence of photon rings becomes less dependent on the latter and more on the former as higher $n$ are considered. Indeed, in the case of the photon rings, the most remarkable property is that the locations $r_n$ ($n>0$) correspond to lower impact parameters compared to the Schwarzschild case, implying a smaller visual appearance of the corresponding bright region on the screen. Furthermore, the tangent of the slope function $\der r_n/\der b$ is an indication of the degree of  demagnification factor of successive photon rings \cite{Perlick:2021aok}. For the non-singular EiBI black hole, the demagnification factor is remarkably higher, meaning a smaller appearance of the back side of the disk than for the standard black hole. We point out that the numerically computed photon rings approach the theoretically calculated critical curve, something expected on fundamental grounds, delimiting the shadow of the black hole in this ray-tracing procedure.

In order to incorporate the accretion disk into this picture, we introduce the emitted specific intensity, $I_{\nu_{\text{e}}}$, at frequency $\nu_{\text{e}}$, such that the observed specific intensity, $I_{\nu_{\text{o}}}$, at frequency $\nu_{\text{o}}$, is related to it through the redshift factor $g=\nu_{\text{o}}/\nu_{\text{e}}$. Due to the conservation of the phase space (in absence of absorption), the quantity $I_\nu/\nu^3$ is conserved along the ray and, therefore, the emitted and observed intensities are related by
\begin{equation} \label{eq:intem}
    I_{\nu_{\text{o}}} = g^3 I_{\nu_{\text{e}}} \ .
\end{equation}
To calculate the redshift factor, we consider a photon with 4-momentum $p_\mu$ that has been emitted from the disk moving with 4-velocity $u^\mu$. The energy of a photon with 4-momentum $p_\mu$ as measured by an observer with 4-velocity $u^\mu$ is given by $\nu = -p_\mu u^\mu$. The redshift factor is then the ratio of the energy measured at the observer's location ($r_o$) to the energy at the emission point ($r_e$): $\nu_o = - (p_\mu u^\mu)|_o$ and $\nu_e = - (p_\mu u^\mu)|_e$, respectively. We assume that both the observer and the emitter are static with respect to the background space-time, meaning we do not consider any orbital motion of the accretion disk. The 4-velocity for a static observer is $u^\mu = u^t \partial_t$, where the time component is determined by the normalization condition to be $u^t = 1/\sqrt{A(r)}$. We place the observer at a large distance from the source ($r_o \gg M$), where the spacetime is asymptotically flat and thus $A(r_o) \approx 1$. The 4-velocities for the observer and emitter are therefore $u^\mu_o = \partial_t$ and $u^\mu_e = (1/\sqrt{A(r_e)}) \partial_t$, respectively. The redshift factor is then given by:
\begin{equation} \label{eq:gg}
    g = \frac{\nu_o}{\nu_e} = \sqrt{A(r_e)}.
\end{equation}
We consider that the emission is monochromatic on the emission frame, $I_{\nu_e} \equiv I(r)$. Therefore, the bolometric intensity is the integral through all the spectra, that is
\begin{equation}
    I^\text{obs} = \int_{\nu_o} I_{\nu_o} \der \nu_o = \int g^3(r)I(r) d \nu_o= g^4 I(r) ,
\end{equation}
where we have combined Eqs.(\ref{eq:intem}) and (\ref{eq:gg}). Finally, we assume that the disk is optically thin and neither absorption nor scattering are relevant when the photon passes through the disk. As a consequence, the observed total intensity for a given ray is the sum of all the emitted intensities, corrected with the redshift factor $g^4$, for which the ray has intersected the plane at $r_m$,
\begin{equation}
    I^\text{obs} = \sum_n g^4 I_{\nu_e}(r_n),
\end{equation}
though for computational purposes we shall consider only up to $n=2$ in view of the demagnification of successive photon rings and the fact that optical thinness cannot be absolute.

\subsection{Semi-analytical emission model}

We describe the local emission intensity of the disk using the models first introduced by Gralla et. al. \cite{Gralla:2020srx}. They employed semi-analytical modeling to fit Standard Unbound-type profiles to the result of simulated General Relativistic MagnetoHydroDynamics (GRMHD) accretion disk images on Kerr space-times with different sets of assumptions for the accreting matter, mimicked in these models via specific picks of parameters. For our system, we used the Palatini approach to obtain the evolution equations. This method does not affect the evolution of matter (which still couples minimally to the geometry), and thus the GRMHD method is similar in EiBI gravity. In addition, we take advantage of the fact that outside the horizon of the non-singular black hole the geometry is basically described by the Schwarzschild space-time plus some small corrections due to Palatini gravity. This way we can reliable employ similar emissions models, suitably adapted to the peculiarities of our gravitational configurations.

The emitted specific intensity employed in this paper is therefore described by the functional form \cite{Gralla:2020srx},
\begin{equation} \label{eq_SUem}
    I(r;\gamma,\mu,\sigma) = \frac{\text{exp}\left( -\frac{1}{2}(\gamma + \text{arcsinh} (\frac{r-\mu}{\sigma}))^2\right)}{\sqrt{(r-\mu)^2+\sigma^2}},
\end{equation}
with three parameters $\{\gamma, \mu, \sigma\}$ that characterize the profile, associated to the profile's rate of growth from asymptotic infinity to its peak, the profile's shift to a given location, and the profile's dilation. For the sake of our analysis we will use the following three sets of parameters:
\begin{align}
    \text{GLM-1:}\quad  & \gamma = - \frac{3}{2}, && \mu=0, \, &\sigma = \frac{M}{2}, \notag\\ 
    \text{GLM-2:}\quad  & \gamma = 0, && \mu=0, & \sigma = \frac{M}{2}, \label{eq:parameters}\\ 
    \text{GLM-3:} \quad  & \gamma = - 2, && \mu=\frac{17}{3}M, & \sigma = \frac{M}{4}, \notag
\end{align}
as suitable adaptations to the spherically symmetric case of previously employed profiles in the rotating (Kerr) case. 
The first two profiles, GLM-1 and GLM-2, characterize emission that monotonically increases approaching the horizon. GLM-3 describes a profile with a peak near the Innermost Stable Circular Orbit (ISCO) for a Schwarzschild black hole. Therefore, the GLM-1/2 and GLM-3 profiles describe different surfaces of peak emission, the former two having different spreads in order to enable visual differences.

\subsection{Face-on imaging}

\begin{figure*}[t!]
    \centering
    \begin{subfigure}{\textwidth}
        \centering
        \includegraphics[width=\linewidth]{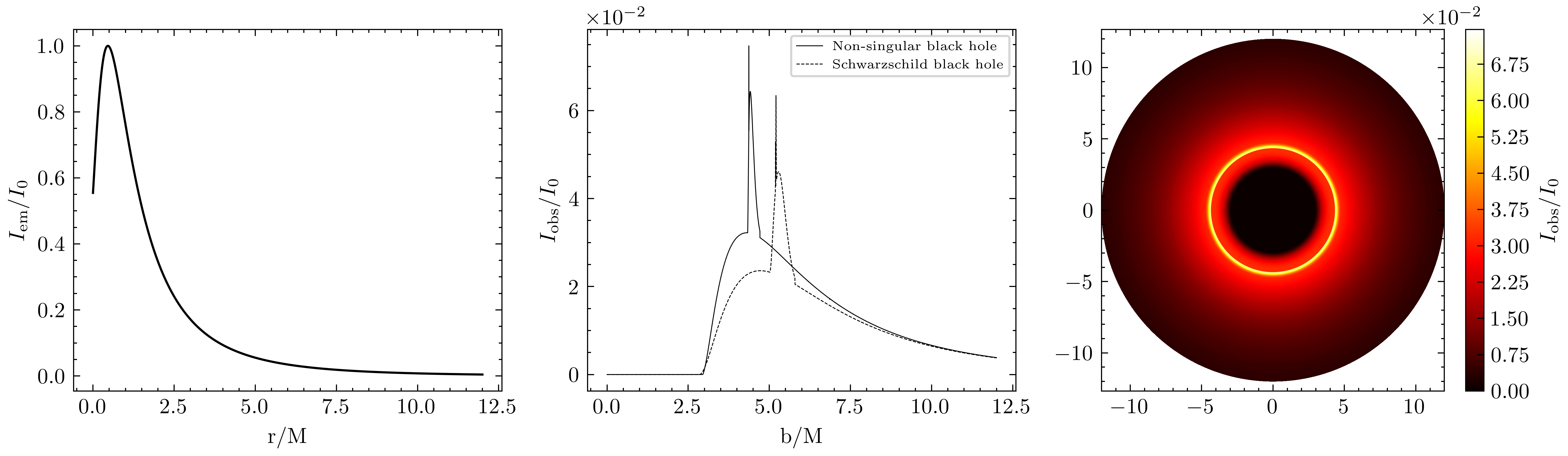} 
        \caption{GLM-1: $\gamma = -3/2$, $\mu = 0$, $\sigma = 1/2$.}
        \label{fig:GLM1}
    \end{subfigure}
    \vfill
    \begin{subfigure}{\textwidth}
        \centering
        \includegraphics[width=\linewidth]{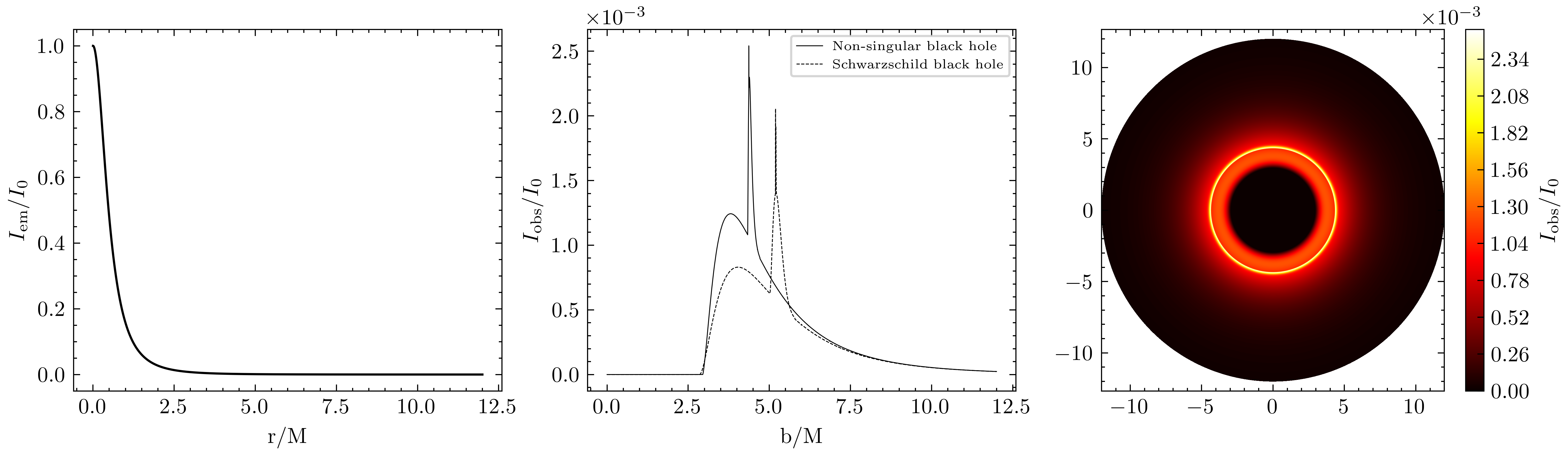} 
        \caption{GLM-2: $\gamma = 0$, $\mu = 0$, $\sigma = 1/2$.}
        \label{fig:GLM2}
    \end{subfigure}
    \vfill
    \begin{subfigure}{\textwidth}
        \centering
        \includegraphics[width=\linewidth]{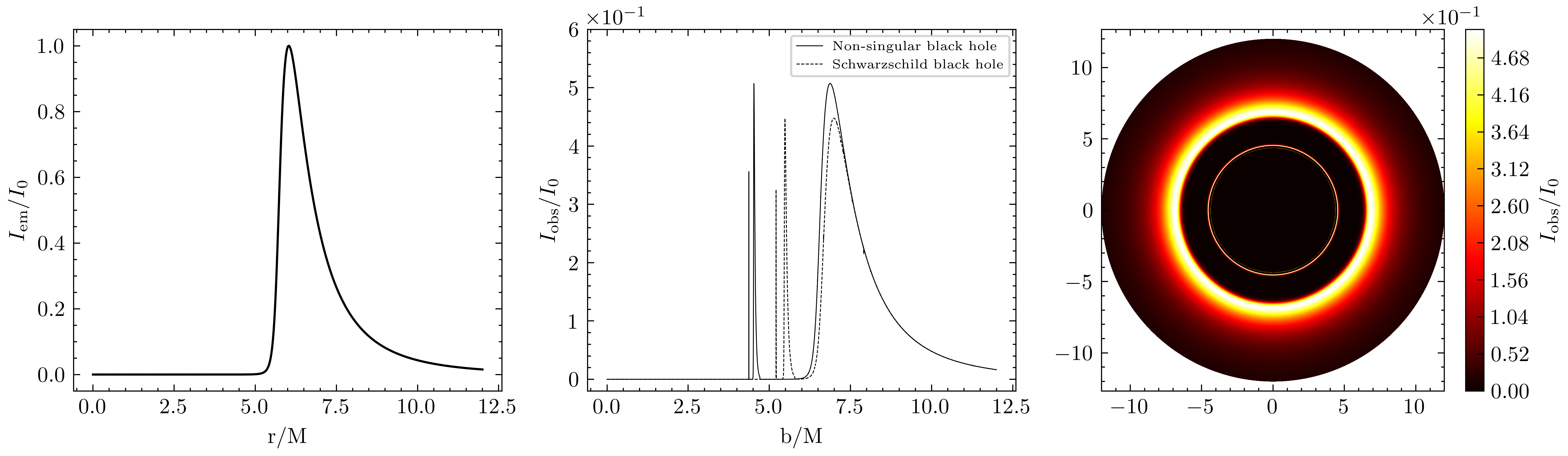} 
        \caption{GLM-3: $\gamma = -2$, $\mu = 17/3$, $\sigma = 1/4$.}
        \label{fig:GLM3}
    \end{subfigure}
    \caption{Optical appearance of a geometrically and optically thin emission disk, with a face-on orientation, near a non-singular EiBI black hole with one horizon supported by an anisotropic fluid. The emitted $I_{em}$ (left plots) and observed $I_{obs}$ (middle plots) intensities are normalized using the maximum value of the emitted intensity $I_{0}$. The emission models employed are the GLM ones, introduced as suitable adaptations of those introduced in \cite{Gralla:2020srx} via Eq.(\ref{eq_SUem}), with the parameters of \eqref{eq:parameters}. The optical appearances plots (right figures) display strong differences in the visibility of their photon rings according to whether the emission extends to the event horizon (top and middle rows) or remains beyond the ISCO (bottom row), and furthermore show moderate differences in the effective region of emission of their photon rings between the non-singular black hole and the Schwarzschild one (solid and dashed curves in the $I_{obs}$ plots.
    }
    \label{fig:Imaging}
\end{figure*}

For the generation of both face-on and inclined (see next section) images, we normalize all the emission models to the peak of intensity of the emitted profile, for a consistent visual comparison. In Fig. \ref{fig:Imaging}, we depict the emitted (left figures) and observed (middle figures) intensities as well as the optical images (right figures), of the three thin-disk GLM models above (from top to bottom) orbiting the non-singular black hole, when viewed face-on (i.e. with a zero inclination between the observer's line of sight with respect to the normal to the disk's surface). In the observed intensity profiles we also depict the corresponding profile for the Schwarzschild black hole as a comparison.

The most prominent characteristic of these images is that they all display a dark central region, where the intensity is strongly dimmed due to both the redshift and much shortened photon's path-length associated to the existence of a horizon at $r \approx 2 M$. Such a central brightness depression appears, in the first two models, well below the critical curve which, as discussed in the previous sections, only determines the theoretical shadow. This is expected as in these disk models light can be emitted from inside the photon sphere, and such photons can escape to infinity, acquiring an important redshift. On the other hand, the central brightness depression very much coincides with the critical curve for the third model, as the emission is far from the photon sphere, near the ISCO, and only the photons that are trapped very nearly the photon sphere are delimiting the critical curve. In this sense, the shadow's size can be approximately taken to be given by the location of the last visible photon ring, as they converge to such a curve at $n \to \infty$. Nonetheless, the shadow in such a third model would still be pierced inside by at least one bright photon rings (i.e. the $n=1$). Qualitatively similar features for these three models have been observed in  other alternative black hole metrics that mildly modify the Schwarzschild black hole, see e.g. \cite{Olmo:2025ctf}.

We also observe that for the three emission models, the sequence of photon rings converge to the theoretically calculated critical curve at $b_c\simeq 4.366 M$ as expected. As with similar black holes which strongly decrease the critical impact parameter, this fact describes a property of the space-time that differs significantly on its optical appearance from the Schwarzschild black hole, where $b_c^\text{Sch} \simeq 5.196M$. The second property is the fact that the redshift applied to the photon during emission is comparatively lower, showing a stronger emission with respect to the Schwarzschild black hole and, consequently, a relatively brighter photon rings as compared to the disk's direct emission, something which manifests via higher peaks in the observed intensity profiles.

For the two first GLM-1/2 models (i.e. those peaking at the horizon), displayed in Figs. \ref{fig:GLM1} and \ref{fig:GLM2}, respectively, the photon rings are overlapped with the disk's direct emission, similarly to what happens for the Schwarzschild black hole at equal emission model. Therefore, for these models the photon rings act as a boost of brightness within a restricted region of the direct emission.  The main difference between the images of these two types of black holes is the fact that the direct emission is much more concentrated for the non-singular black as compared to the much more spread emission from the Schwarzschild one. This is a consequence of the form of the metric function in Eq.\eqref{fig:metricomp}, that falls very sharply near the horizon of the black hole.

The third GLM-3 model's optical appearance, displayed in Fig. \ref{fig:GLM3}, describes a more astrophysically realistic accretion disk where the matter is orbiting the region outside the ISCO. This case is the most feature-rich of them all, showing the characteristic properties of the non-singular black hole. The first of them is the overall lower redshift affecting the photons. The second one is the inner position of the photon rings compared to the Schwarzschild black hole, yielding a noticeable separation from the direct emission. And the third one is that the separation between the photon rings is shorter, giving a closer appearance of such rings. These properties yield a neat visual difference between both black holes: the non-singular black hole has a smaller ring diameter showing a bigger separation to the direct emission.

\subsection{Inclined imaging}

\begin{figure*}[t!]
    \centering
\includegraphics[width=0.95\textwidth]{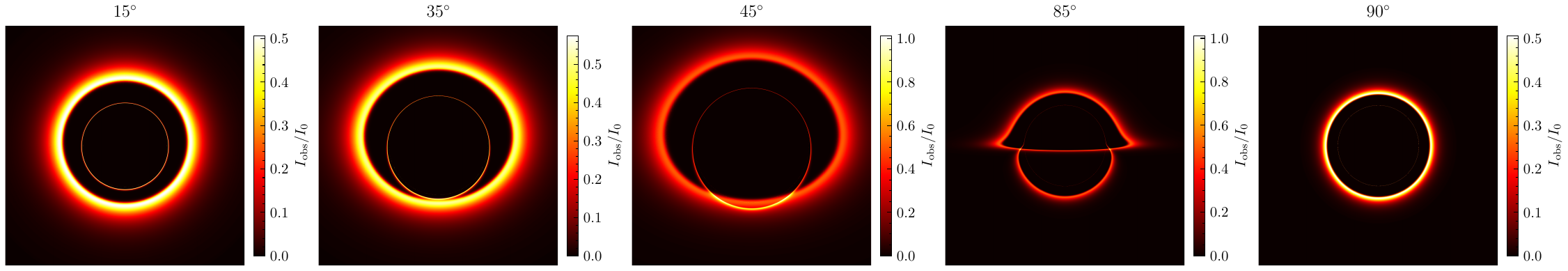}
    \centering
    \captionsetup{justification=centering}
    \caption*{Non-singular Palatini black hole} 
    \includegraphics[width=0.95\textwidth]{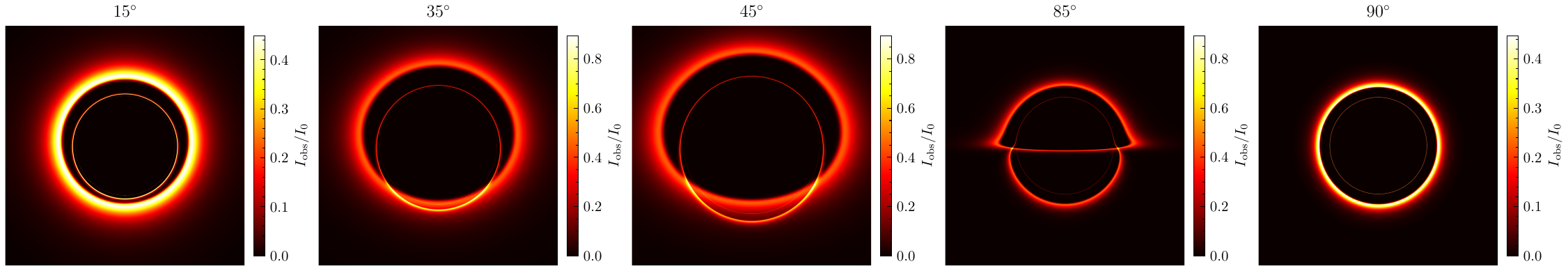} %
    \captionsetup{justification=centering}
    \caption*{Schwarzschild black hole}
    \caption{Optical appearance of the geometrically and optically thin emission with inclination angles of (from left to right) $15^\circ$, 35$^\circ$, $45^\circ$, $85^\circ$ and $90^\circ$ with respect to the normal direction of the disk, for the non-singular EiBI (top) and Schwarzschild (bottom) black holes. The images are in $26 \times 26$ units of mass of each black hole.}
    \label{fig:inclined}
\end{figure*}

Accretion disks with emission originating beyond the ISCO represent a physically plausible and robust model for astrophysical accretion in equilibrium. While a near face-on disk might be observed in specific (and lucky) cases, in general it is more likely to encounter disks with some degree of inclination relative to our line of sight (as the current observations of M87). Considering these facts, comprehending the visual appearance of an inclined disk is important to interpret astronomical data. To understand the unique characteristics of the black hole in these scenarios, we numerically simulate and plot images of different inclinations of the non-singular black hole with an ISCO-dominated GLM-3 model, and compare them with the standard Schwarzschild black hole. The corresponding results are displayed in Fig. \ref{fig:inclined}, viewed at the inclinations $15^\circ$, 35$^\circ$, $45^\circ$, $85^\circ$ and $90^\circ$. The distinctive characteristics previously discussed for the face-on orientation persist under inclination, albeit with  subtle but significant changes for the observed image, summarized as follows:
\begin{itemize}
    \item The central brightness depression and the surrounding photon rings from the non-singular black hole appear smaller in angular size than those from the Schwarzschild black hole. This is a consequence of the differences between the corresponding metrics, and it is preserved across all inclinations.
    \item With increasing inclination, the photon rings suffer a slight distortion, appearing to be asymmetric in their width, due the repeated image of the disk being distorted from the perspective. This effect is compounded by the intrinsically thinner photon rings in the non-singular black holes due to the reduced location of the photon sphere. As a direct consequence, resolving distinct higher-order photon rings for the non-singular black hole demands significantly higher angular resolution as compared to images of a Schwarzschild black hole.
    \item A critical observational implication of the smaller photon rings of the non-singular black hole is its effect on the overlap with the direct emission of the disk. For the Schwarzschild case, the photon ring starts to visually overlap with the direct disk emission at a relatively moderate inclination, as we show in the 15$^\circ$ inclination plot. In contrast, the photon ring of the non-singular black hole does not experience such an overlap until significantly higher inclination is reached, as shown for 35$^\circ$ inclination. While this distinction may be less critical for optically thin disks, this difference is critical for thick disk: it dictates whether the full longitude of the photon rings remain resolvable or become outshined  by the direct emission from the disk.
\end{itemize}

\section{Observing the Lyapunov exponent} \label{S:V}
After finding and discussing the full optical images, we now make use of the observational opportunities present in the lensing Lyapunov exponent, resuming our discussion of Sec. \ref{sec:lya}. Let us recall such an exponent can be calculated using the relation of angular deflection experienced by two photons, and the scaling relation presented in \cite{Kocherlakota:2024hyq}. For an extended emission located at the equator, the width $w_n$ of subsequent photon rings are related, in the $n\to \infty$ limit, as (in face-on orientation) \cite{Johnson:2020eaaz1310}
\begin{equation}
    \frac{w_{n+1}}{w_n} \simeq e^{-\gamma_{ps}}. \label{eq:width_ratio}
\end{equation}
Therefore, measuring the width of the peaks of emission related to the two first photon rings yield an approximate value of the Lyapunov lensing index. The validity of this procedure comes from the fact that photon rings approach very quickly the critical curve even for low values of $n$, so the information provided by the two first photon rings is expected to give an accurate estimate of the Lyapunov index. It should be pointed out that there is an alternative method of measuring such an exponent consisting on using the flux ratio, the latter obeying the same rule, that is
\begin{equation}
    \frac{I_{n+1}}{I_n} \simeq e^{-\gamma_{ps}} \label{eq:width_ratio} ,
\end{equation}
where the photon ring's flux $I_n$ can be found via integration of the ``area" spanned below the emission peak. This latter method has the drawback of depending on how ``concentrated" the emission profile is around its peak, since photons turning a number of half-turns will see different emission regions, and therefore without a full knowledge of the features of the emission region any measurement of the Lyapunov exponent is prone to significant errors (see e.g. \cite{daSilva:2023jxa} for a discussion on this point). On the other hand, both the observed flux and the width methods have the drawback that the EHT Collaboration cannot measure brightness contrasts below a certain threshold, given by a $\sim 10\%$ of the peak peak image brightness  \cite{EventHorizonTelescope:2022xqj}. For the sake of this paper we opt for using the width method.

\begin{figure}
    \centering
    \includegraphics[width=0.9\linewidth]{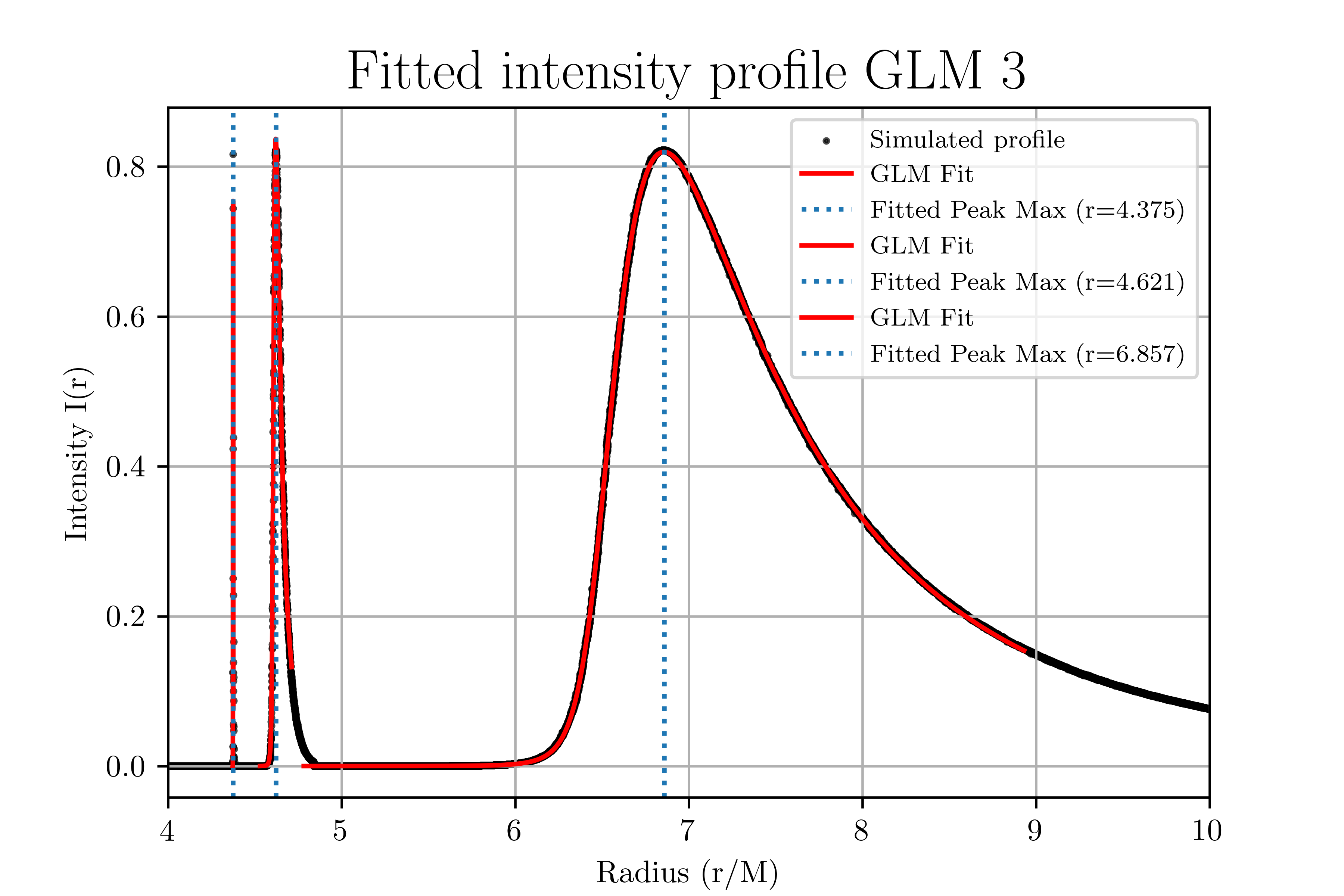}
    \caption{Main peaks of emission fitted to the GLM 3 profile}
    \label{fig:fitted_peaks}
\end{figure}

For our purposes we use the GLM-3 profile as their photon rings can be resolved from the direct emission, and, as we discussed before, because it is a plausible emission profile. Additionally, as we are taking a ratio between widths of photon rings, we suppress most of the uncertainty associated to different models for the matter orbiting in the disk. From the observed intensity profile, $I(r)$, we detect the three peaks associated to the direct emission and the first two photon rings, and fit them using the GLM distribution. We plot the fitted peaks in Fig. \ref{fig:fitted_peaks} and the parameters of the fit for each peak in Table \ref{tab:glm_fit_peaks}.

\begin{table}[t!]
\centering
\begin{tabular}{rcccc}
\toprule
&Peak ($\mu/M$)  & $\gamma$ & $\sigma$ & FWHM \\
\midrule
2nd Photon Ring & 4.374 & -1.5 & $2.4 \times 10^{-4}$                 & $9.6 \times 10^{-4}$ \\
1st Photon Ring & 4.605 & -1.8 & $1.3 \times 10^{-2}$ & $5.1 \times 10^{-2}$ \\
Direct Emission & 6.450 & -2.1 & 0.25                 & 1.2                  \\
\bottomrule
\end{tabular}
\caption{GLM fit parameters for observed intensity peaks.}
\label{tab:glm_fit_peaks}
\end{table}

In order to calculate the width of the peaks, we use the Full Width at Half Maximum (FWHM) for each peak. This method provides a quantitative measure of how broad a given feature (in our case the thickness of the ring) appears in an image and, specifically, it works by taking the width of the peak as given by the locations in the impact space parameter for which the intensity drops to half of its maximum value. Subsequently, we use the width ratio \eqref{eq:width_ratio} with the peaks associated to the two photon rings, to estimate the Lyapunov lensing exponent. This method yields the result
\begin{equation}
    \gamma_{\text{ps;width}} \approx 3.2.
\end{equation}
This is a relative deviation of $\sim 8\%$ as compared to the theoretical value calculated in \eqref{eq:lyapunov_th}. If we recall that the flux ratio is an exponential version of the Lyapunov exponent, this represents a significant difference of $\sim 25\%$ between the theoretical quantity and the observational one via the width method. This deviation is thus larger than the one  assuming a Schwarzschild black hole, whose lensing Lyapunov was given by $\gamma^\text{Sch} = \pi$. It is worth pointing out that, while this is certainly quite an important difference with respect to the theoretically-computed number, it is of the same order than the deviations found then such estimations are made following the observed flux method; for instance, in \cite{Olmo:2023lil} a similar configuration, EiBI-supported, was considered with the result that the \textcolor{blue}{flux} ratio between the $n=1$ and $n=2$ photon rings overestimate such a ratio of the order of the $\sim 20\%$ for a GLM-3 model (see Tables III and IV in that paper) due to the unknowns in the disk's modeling. In fact, it has been recently pointed out in \cite{Urso:2025gos} that, for parametrized black hole solutions, using the rings' width as a proxy for modifications to the background geometry is typically drowned in the disk's emission features (such as its fall-off), at least within the current interferometric capabilities. All together, these results indicate that the measurable quantities from the unstable photon sphere around the non-singular EiBI black hole are difficult to distinguish from the ones around a Schwarzschild black hole.

On the other hand, using Eq.  \eqref{eq:lyapunov_time}, we can estimate the characteristic coordinate time of the instabilities around the photon sphere using the values observed in the simulation, yielding
\begin{equation}
    t_\text{ps;width} \simeq 4.5 M.
\end{equation}
This result is mildly different from the theoretically-computed one of Eq.(\ref{eq:lyapunov_thh}) and can be furthermore compared with the one of the Schwarzschild black hole case $t_\text{ps;Sch} \approx 5.196M$
yielding a moderately smaller value of $\sim 13\%$ for the instability time-scale of a null geodesic nearing the photon sphere. This hints that differences between both solutions could be easier to be detected via hot-spot observations.

The persistent difficulties in reconciling theoretical quantities related to unstable bound orbits, the phenomenological methods  via the set of Lyapunov quantities and their width and observed flux methods, the unknowns in the disk's modeling, and the current limitations to carry out precise observations of the features of the photon rings, makes a case for the search for alternative methods, such as the photon autocorrelations proposed in  \cite{Hadar:2020fda}, the relative separation between neighboring photon rings \cite{Aratore:2024bro}, or via their overlapping \cite{Tsupko:2025hhf}. In particular, the correspondence between quasi-normal modes and black hole imaging hinted in Eq.(\ref{eq:corre}) could provide some help despite the fact that its range of applicability is associated to large $l$; nonetheless errors in the correspondence for smaller values of $l$ could be acceptable for providing useful information on black hole imaging should they be of the same order of those associated to the flux/width methods. 

\section{Conclusion} \label{S:VI}

In this work, we have investigated the optical appearance of a non-singular black hole arising from Eddington-inspired Born-Infeld gravity coupled to an anisotropic fluid, corresponding to the pool of solutions found in \cite{Menchon:2017qed}. Our primary goal was to determine if this non-singular black hole model produces observable signatures on its optical appearance that would distinguish it from a standard Schwarzschild black hole. To this end we picked a combination of parameters selecting a black hole with a single horizon located very approximately at the same Schwarzschild radius, and focused on characterizing the features of its photon rings and the central brightness depression in comparison to the those of the Schwarzschild one.

We began by deriving the critical parameters of the space-time, in particular the radius of the photon sphere due to its tight connection to both photon rings and shadow features of the image. Subsequently, we developed a numerical ray-tracing approach to generate images of the black hole illuminated by both a face-on and inclined accretion disk. These simulations delivered a critical difference between both black holes: the smaller size of shadow (defined this way by the inner region to the critical curve on the observer's screen) from the non-singular black hole as compared to the Schwarszchild one. Furthermore, we hinted that such a reduced size shadow would also anticipate relevant differences in the structure of both photon rings and the actual size of the central brightness depression.

In order to consolidate this expectation, a central point to our analysis is the characterization of the lensing Lyapunov exponent driving the radial and temporal drift of nearly-bound unstable geodesics, which are relevant for time-averaged images and transient hot-spots events, respectively. In fact, successive photon rings (provided the disk is not spherically symmetric and has sufficiently large optical depth) obey an exponential rule in the ratio of their widths and luminosities driven by this Lyapunov exponent, which provides a chance for their observational characterization.

To implement this, we considered suitable adaptations of the Standard Unbound distribution previously employed in the literature to match specific results of GRMHD simulations, via the three GLM models first introduced by Gralla, Lupsasca and Marrone. We generated both face-on and inclined images (at several angles) of these three models, two of which represent emission profiles reaching up to the event horizon, and one having its emission outside the ISCO. The overall images display significant differences as compared to the Schwarzschild case, both in terms of a reduction in the size of the central brightness depression, and in relatively brighter photon rings, also displaced in their positions in the image. These features are moderately modified when considering inclined images. Furthermore, by fitting the intensity profile of the rings from our simulated images and using the Full Width at Half Maximum method, we obtained a numerical estimation of the lensing Lyapunov lensing exponent $\gamma_{ps}$, that quantifies the instability of the orbit. Such an estimation deviates a $\sim 8\%$ from the theoretical expectation, and therefore a $\sim 25\%$ on its exponential version, which controls the actual flux ratio decay between successive photon rings. This way, this quantifier falls within the range of the Lyapunov exponent for the Schwarzschild solution. This means that it would not be possible to distinguish between the non-singular EiBI black holes studied in this work and the Schwarzschild one with this method alone, though better opportunities could be present for the Lyapunov time, given the fact that it presents a moderate decrease as compared to the expectations of Schwarzschild geometry.

Our results show an important observational degeneracy: this non-singular black hole model, despite its regular center, and having its horizon very approximately located at the Schwarzschild radius but a photon sphere at a significantly modified location, produces static images that are qualitatively similar to those of a Schwarzschild black hole, yet with different quantitative features, mainly related to the size of the central brightness depression and to the photon rings' distance. However, distinguishing these black holes based on the structure of the first two photon rings remains a challenge for current imaging techniques. It is relevant to emphasize that the current limitations of the width/flux approach may, at least in part, account for its reduced sensitivity, hampering our ability to clearly distinguish between the Schwarzschild black hole and the solutions derived from the EiBI gravity analyzed in this study. In particular, while the equatorial approximation proves useful, it may be overly simplistic. This suggests that contributions from higher latitudes could be significant, and that such an oversimplification may partly explain the challenges in achieving a meaningful comparison with the Schwarzschild case. Future work should therefore explore dynamical phenomena within the framework of Palatini gravity. Furthermore, observational channels such as the analysis of hot-spots or the ringdown phase of gravitational waves from black hole mergers offer a promising avenue to break this degeneracy and probe the nature of space-time in the strong-field regime. Work along these lines is currently underway.

\section*{Acknowledgements}

A.~R.  would like to express his gratitude to Silesian University in Opava, Czech Republic, for their financial support, and is very grateful for the hospitality of the University of Valencia and the Complutense University of Madrid (Spain). The creation of this article was supported by the grant program Vouchers for Universities in the Moravian-Silesian Region (registration number CZ.10.03.01/00/23\_042/0000390).
This work is supported by the Spanish National Grants PID2022-138607NBI00 and CNS2024-154444, funded by MICIU/AEI/10.13039/501100011033 (Spain).

\appendix 

\section{Derivation of spherically symmetric solutions in Palatini gravity with an anisotropic fluid}  \label{App1}

We summarize here the main ingredients behind the theoretical setting generating the line element (\ref{eq:lineel}), and corresponding to the results found in \cite{Menchon:2017qed}, suitably cast for the purposes of the present work. The action is split as $\mathcal{S}=\mathcal{S}_g + \mathcal{S}_m$ where the gravitational sector $\mathcal{S}_g$ corresponds to
\begin{equation} \label{eq:actionEiBI}
\mathcal{S}_g=\frac{1}{\kappa^2 \epsilon} \int d^4x  \left(\sqrt{ \vert g_{\mu\nu} + \epsilon R_{\mu\nu} \vert}-\sqrt{\vert g_{\mu\nu} \vert} \right),
\end{equation}
where vertical bars denote a determinant, $R_{\mu\nu}$ is the symmetric part of the Ricci tensor, and $\epsilon$ is a parameter with dimensions of length squared. In the regime $\epsilon \ll \vert R_{\mu\nu} \vert$ the above gravitational action recovers the Einstein-Hilbert one of GR plus quadratic curvature corrections. This is the so-called Eddington-inspired Born-Infeld (EiBI) gravity, a viable modification of GR and with many different applications, see \cite{BeltranJimenez:2017doy} for a review.  

The matter side employed to generate (\ref{eq:lineel}) is given by an anisotropic fluid with energy-momentum tensor
\begin{equation} \label{eq:mattertmunu}
    {T_\mu}^{\nu}= \text{diag}(-\rho,-\rho,\rho+s\beta^4 \rho^2,\rho+ s\beta^4 \rho^2),
\end{equation}
where $\beta$ is a parameter characterizing the fluid. The sign $s=\pm 1$ of $\beta$ furthermore controls whether the energy density of the fluid is finite (for $s=+1$) or divergent (for $s=-1$). Among the recognized different matter sources encoded within this fluid energy-momentum tensor are those of non-linear electrodynamics. In the latter case, the kind of solutions given by  (\ref{eq:lineel}) may be also be generated using quadratic gravity  due to the coincidental structure of their field equations with those of EiBI gravity  for non-linear electrodynamics \cite{Afonso:2018mxn}.

A key ingredient in this setting is the employ of the Palatini formalism: independent variation of the action with respect to metric and connection. The corresponding two sets of equations can be solved using different strategies developed in past works, arriving to the line element (\ref{eq:lineel}). Such a line element is generally parametrized in terms of two scales, associated to the gravity and matter sector as follows:
\begin{itemize}
\item Gravity: The EiBI parameter is recast as $\epsilon=-l_{\epsilon}^2$, where $l_{\epsilon}^2$ is the gravity's length squared, in agreement with the units of the action (\ref{eq:actionEiBI}).
\item Matter: we introduce a new length scale as $l_m^2=\frac{\vert \beta \vert}{2}$, associated to the energy-momentum tensor (\ref{eq:mattertmunu}).
\end{itemize}
We also introduce introduce the variable $r_c=r_0 \rho_0^{1/4} \vert \beta \vert$ (to deal with radial coordinates $z=r/r_c$), where $r_0$ is an integration constant and $\rho_0$ another integration constant related to the fluid's density and its conservation equation. For the sake of this work we fix $r_0=\rho_0=1$ so that $r_c= \vert \beta \vert$. Furthermore, the combination of the signs of $\epsilon$ and $\beta$ creates four types of solutions with very different behaviours among each other, such as regular and singular black holes, traversable wormholes, and de Sitter cores, all of them satisfying the energy conditions, as described in \cite{Menchon:2017qed}. For the sake of this work we took the choice of $\epsilon<0$ and $\beta>0$ (Type A solutions), corresponding to the specific functions defined below Eq.(\ref{eq:lineel}) and which describe non-singular black holes according to the criterion of geodesic completeness.

\section{Numerical calculation of the Appell functions} \label{App2}

The metric functions for non-singular black hole spacetime \eqref{eq:lineel} are expressed using the Appell F1 hypergeometric function in its analytical form,
\begin{equation} \label{eq:AppellF1}
    F_1 \left( \alpha, \beta_1, \beta_2, \gamma; x,y \right) = \sum_{m=0}^\infty \sum_{n=0}^\infty \frac{(\alpha)_{m+n} ( \beta_1)_m (\beta_2)_n}{m! n! (\gamma)_{m+n}} x^m y^n, 
\end{equation}
with the Pochhammer symbols,
\begin{equation}
    (z)_n \equiv \frac{\Gamma(z+n)}{\Gamma(z)},
\end{equation}
expressed using the Gamma functions,
\begin{equation}
    \Gamma(z) \equiv \int_0^\infty t^{z-1} e^{-t} \der t.
\end{equation}

Most of the symbolic calculation software, such as \textit{Mathematic\text{@}}, have some implementation of the hypergeometric function such as the Appell function \eqref{eq:AppellF1}. However, when numerically solving the null geodesic equation \eqref{eq:geodesic}, we find that such analytical implementations result in not manageable times of integration, as the calculation of the Appell functions goes to unneeded precision and redundant calculations for the Pochhammer symbols. Therefore, we implemented our own numerical method to calculate the Appell F1 function $F_1 \left( \alpha, \beta_1, \beta_2, \gamma; x,y \right)$ that goes as follows,

\begin{enumerate}
    \item We take a finite limit to the sum of the $m$ and $n$ indices, to $m_\text{max}$ and  $n_\text{max}$, considering that they will converge to a beforehand specified precision, $$ F_1 \left( \alpha, \beta_1, \beta_2, \gamma; x,y \right) = \sum_{m=0}^{m_\text{max}} \sum_{n=0}^{n_\text{max}} \left(\dots\right).$$
    \item Before doing the sums of the Appell function, precalculate the powers $x^m$ and $y^m$, avoiding redundant calculations. This reduces the operations from $\mathcal{O}(m_\text{max}\cdot n_\text{max})$ to $\mathcal{O}\left(m_\text{max} + n_\text{max}\right)$.
    \item Before doing the sums of the Appell function, precalculate the Pochhammer symbols, $(\alpha)_{m+n}$, as the sum $m+n$ yields redundant values when performing the sums. We perform the same method for $(\gamma)_{m+n}$, and the Gamma functions $\Gamma(m+1)$ and $\Gamma(n+1)$. For our specific metric, $\beta_1 = \beta_2$, so we can also precalculate the values of $(\beta_1)_m$, and share them with $(\beta_2)_n$. Each of these methods reduces the calculations from $\mathcal{O}\left(m_\text{max} \cdot n_\text{max}\right)$ to $\mathcal{O}\left( m_\text{max} + n_\text{max} \right)$.
    \item Calculate the sum by checking the precalculated values of the Pochhammer and Gamma functions. We add a convergence check to avoid summing extra values below our precision.
\end{enumerate}
Following this method, we go from the brute-force algorithm of order $\mathcal{O}(N^2)$ to order $\mathcal{O}(2N)$, being $N$ the number of terms in the sum. We have found that $m_\text{max}= n_\text{max} = 10$ converges the integral to $10^{-5}$ precision in manageable times.

With this numerical attack of the hypergeometrical functions of our metric, we ray-trace the geometry (\ref{eq:lineel}) with assumptions for the disk's features stated on the main text, via a suitable implementation of our GRAVITYp code based on \textit{Mathematic\text{@}} and Python.

\end{document}